\documentclass{article}

\usepackage{arxiv}

\usepackage[utf8]{inputenc} 
\usepackage[T1]{fontenc}    
\usepackage{hyperref}       
\usepackage{url}            
\usepackage{booktabs}       
\usepackage{amsfonts}       
\usepackage{nicefrac}       
\usepackage{microtype}      
\usepackage{lipsum}		
\usepackage{graphicx}
\usepackage{doi}
\usepackage{multirow}
\usepackage{subfig}
\def\BibTeX{{\rm B\kern-.05em{\sc i\kern-.025em b}\kern-.08em
    T\kern-.1667em\lower.7ex\hbox{E}\kern-.125emX}}

\title{Ultrasound Detection of Subquadricipital Recess Distension}

\author{ \href{https://orcid.org/0000-0002-7445-7615}{\includegraphics[scale=0.06]{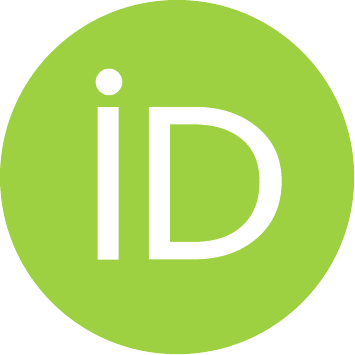}\hspace{1mm}Marco Colussi$^1$}\\
	\texttt{marco.colussi@unimi.it} \\
    \And
	\href{https://orcid.org/0000-0002-8247-2524}{\includegraphics[scale=0.06]{orcid.pdf}\hspace{1mm}Gabriele Civitarese$^1$}\\
	\texttt{gabriele.civitarese@unimi.it} \\
	\And
	\href{https://orcid.org/0000-0001-5745-1230}{\includegraphics[scale=0.06]{orcid.pdf}\hspace{1mm}Dragan Ahmetovic$^1$}\\
    \texttt{dragan.ahmetovic@unimi.it} \\
	\And
	\href{https://orcid.org/0000-0002-1727-7650}{\includegraphics[scale=0.06]{orcid.pdf}\hspace{1mm}Claudio Bettini$^1$}\\
    \texttt{claudio.bettini@unimi.it} \\
	\And
	\href{https://orcid.org/0000-0001-6465-7624}{\includegraphics[scale=0.06]{orcid.pdf}\hspace{1mm}Roberta Gualtierotti$^{2,3}$}\\
    \texttt{roberta.gualtierotti@unimi.it}\\
    \And
    \href{https://orcid.org/0000-0001-7423-9864}{\includegraphics[scale=0.06]{orcid.pdf}\hspace{1mm}Flora Peyvandi$^{2,3}$}\\
    \texttt{flora.peyvandi@unimi.it}
    \And
    \href{https://orcid.org/0000-0002-8416-4023}{\includegraphics[scale=0.06]{orcid.pdf}\hspace{1mm}Sergio Mascetti$^1$}\\
    \texttt{sergio.mascetti@unimi.it} \\
    }
\date{%
    $^1$Università degli studi di Milano, Dipartimento di Informatica\\
    Via Celoria, 18 20133, Milan Italy\\%
    $^2$Università degli Studi di Milano, Dipertimento di Fisiopatologia Medico-chirurgica e dei Trapianti\\
    Via Pace, 9, 20122, Milan, Italy\\[2ex]%
    $^3$Fondazione IRCCS Ca’ Granda, Ospedale Maggiore Policlinico di Milano, \\
    Centro Emofilia e Trombosi, Angelo Bianchi Bonomi\\
    Via Pace, 9, 20122, Milan, Italy\\%
}

\renewcommand{\shorttitle}{Ultrasound Detection of Subquadricipital Recess Distension}

\hypersetup{
pdftitle={Ultrasound Detection of Subquadricipital Recess Distension},
pdfsubject={q-bio.NC, q-bio.QM},
pdfauthor={Marco Colussi},
pdfkeywords={multi-task learning, clinical decision support, ecography, hemarthrosis},
}

\begin{document}

\maketitle

\begin{abstract}
	Joint bleeding is a common condition for people with hemophilia and, if untreated, can result in hemophilic arthropathy. Ultrasound imaging has recently emerged as an effective tool to diagnose joint recess distension caused by joint bleeding. However, no computer-aided diagnosis tool exists to support the practitioner in the diagnosis process. 
This paper addresses the problem of automatically detecting the recess and assessing whether it is distended in knee ultrasound images collected in patients with hemophilia.
After framing the problem, we propose two different approaches: the first one adopts a one-stage object detection algorithm, while the second one is a multi-task approach with a classification and a detection branch.
The experimental evaluation, conducted with $483$ annotated images, shows that the solution based on object detection alone has a balanced accuracy score of $0.74$ with a mean IoU value of $0.66$, while the multi-task approach has a higher balanced accuracy value ($0.78$) at the cost of a slightly lower mean IoU value. 
\end{abstract}


\section{Introduction}
Hemophilia is a hereditary blood coagulation disorder that results in an increased risk of bleeding, due to trauma or spontaneously, which worsens with the severity of the disease.
Bleedings can frequently occur also inside joints (mostly ankles, knees and elbows) and muscles, which together account for around $80\%$ of the bleeding events in patients with Hemophilia~\cite{ROOSENDAALGoris2003Bjdi,srivastava2020wfh}. 
Joint bleeding causes joint recess \textit{distension} which, if recurrent, can result in synovial hyperplasia, osteochondral damage,
and hemophilic arthropathy~\cite{HilgartnerMargaretW2002Ctoh}.
Thus, it is essential to promptly recognize joint recess distension.

Physical examination may not be sufficient to diagnose joint recess distension, since in the early stage it can be asymptomatic~\cite{PlutDomen2019Daoh}.
Magnetic Resonance Imagining (MRI) is generally considered the gold standard tool for precise evaluation of joints but it is not practical for regular follow-up of patients with haemophilia due to the high costs, limited availability and long examination times~\cite{PlutDomen2019Daoh}.
An alternative solution is ultrasound (US) imaging \cite{WellsPNT2006Ui} that, contrary to MRI, has a low cost, short examining time and it is widely accessible \cite{JOSHUAFredrick2007Sfoa}.
\textit{Hemophilia Early Arthropathy Detection with UltraSound} (HEAD-US) is a standardized protocol designed to guide the practitioner in acquiring relevant US images and interpreting them for the diagnosis of joint recess distension in the $6$ most commonly affected joints~\cite{MartinoliCarlo2013Dado}.

Computer aided diagnosis (CAD) systems can improve detection accuracy~\cite{chan1990improvement} and reduce the image reading time required by the practitioners~\cite{doi2005current}.
The potential effectiveness of US-based CAD systems to support the diagnosis of joint distension in people with hemophilia is suggested by recent studies that focus in identifying joint effusions related to injuries~\cite{long2020segmentation}. 

In this work, we formulate the research problem
of supporting the physicians in diagnosing joint recess distension in patients with hemophilia using a CAD system.
The problem consists of detecting the joint recess inside US images and classifying it as \textit{Distended} or \textit{Non-distended}.
Specifically, we focus on the main joint recess of the knee, also called \textit{SubQuadricipital Recess} (SQR). We consider the SQR longitudinal scan, which is one of the three scans specified in HEAD-US protocol for this joint~\cite{MartinoliCarlo2013Dado}.
One prior work addresses the problem of detecting SQR distension in pediatric patients with hemophilia~\cite{tyrrell2021detection}, but specific details about the methodology and the evaluation are not reported.

Besides formulating the research problem, we also propose two approaches to address it.
The first one, called the \textit{Detection approach}, adopts state-of-the-art object detection to find \textit{Distended} or \textit{Non-distended} SQRs inside the US image and returns the detection having the highest confidence.
The second solution, called the \textit{Multi-task approach} uses a multi-task learning process, with the aim of simultaneously detecting the SQR inside the US image and classifying it as \textit{Distended} or \textit{Non-distended}.

The experiments, conducted on a dataset of images we collected and annotated from $200$ adults subjects with hemophilia, reveal that the \textit{Multi-task approach} improves over the \textit{Detection approach} in terms of classification accuracy.
Indeed, the Balanced Accuracy (BA) is $0.78$ with the \textit{Multi-task approach} and $0.74$ with the \textit{Detection approach}. In particular, sensitivity improves from $0,52$ with the \textit{Detection approach} to $0,64$ with the \textit{Multi-task approach}.
Instead, for what concerns the detection accuracy, the \textit{Detection approach} has slightly better performance than the \textit{Multi-task approach}:
the average Intersection over Union (IoU) is $0.66$ for the  \textit{Detection approach} and $0.63$ for the \textit{Multi-task approach}.

To sum up, the novel contributions of this paper are the following:
\begin{itemize}
    \item We formulate the problem of detecting and classifying distensions in SQR from US images.
    \item We propose two solutions to tackle this problem.
    \item We evaluate and compare the proposed solutions on a dataset collected from $200$ patients.
\end{itemize}

\section{Problem formulation}\label{sec:prob}
In this research, we address the problem of the automated detection of the subquadricipital recess (SQR) and its classification as \textit{Distended} or \textit{Non-distended}.

\subsection{Ultrasound images}
Ultrasound (US)~\cite{chan2011basics} is a very popular medical imaging technique.
It is portable, safe and affordable and therefore commonly used in healthcare~\cite{brattain2018machine}.
However, some limitations of this technique are the high dependence on the operator expertise level and possible noisiness of the acquired images~\cite{PlutDomen2019Daoh}.

US imaging uses sound waves at high frequencies. The reflections of the signal are then measured to represent the image. This technique can produce images with a high spatial resolution of the internal structures of the body, like tendons, bones, blood, and muscles~\cite{WellsPNT2006Ui}. The images are represented in grayscale, where each pixel value describes the density of the material the signal encounters. Light areas represent echogenic tissues (\textit{i.e.}, that reflect sound waves) like bones, while dark areas represent anechogenic (\textit{i.e.}, that do not reflect sound) structures such as liquids.
Another effect to take into account is that echogenic tissues, such as bone, shield the signal that is unable to travel through them, thus making it impossible to detect anything below them.
An example is shown in Figure~\ref{fig:probe_pos}: the patella is clearly distinguishable in light color (see the red box) while the area below it is almost completely black.


\subsection{Subquadricipital Recess longitudinal scan}\label{sec:sqr}

\begin{figure}[h!]
\centering
\subfloat[Example of SQR longitudinal scan\label{fig:example}]{
\includegraphics[width=.4\textwidth]{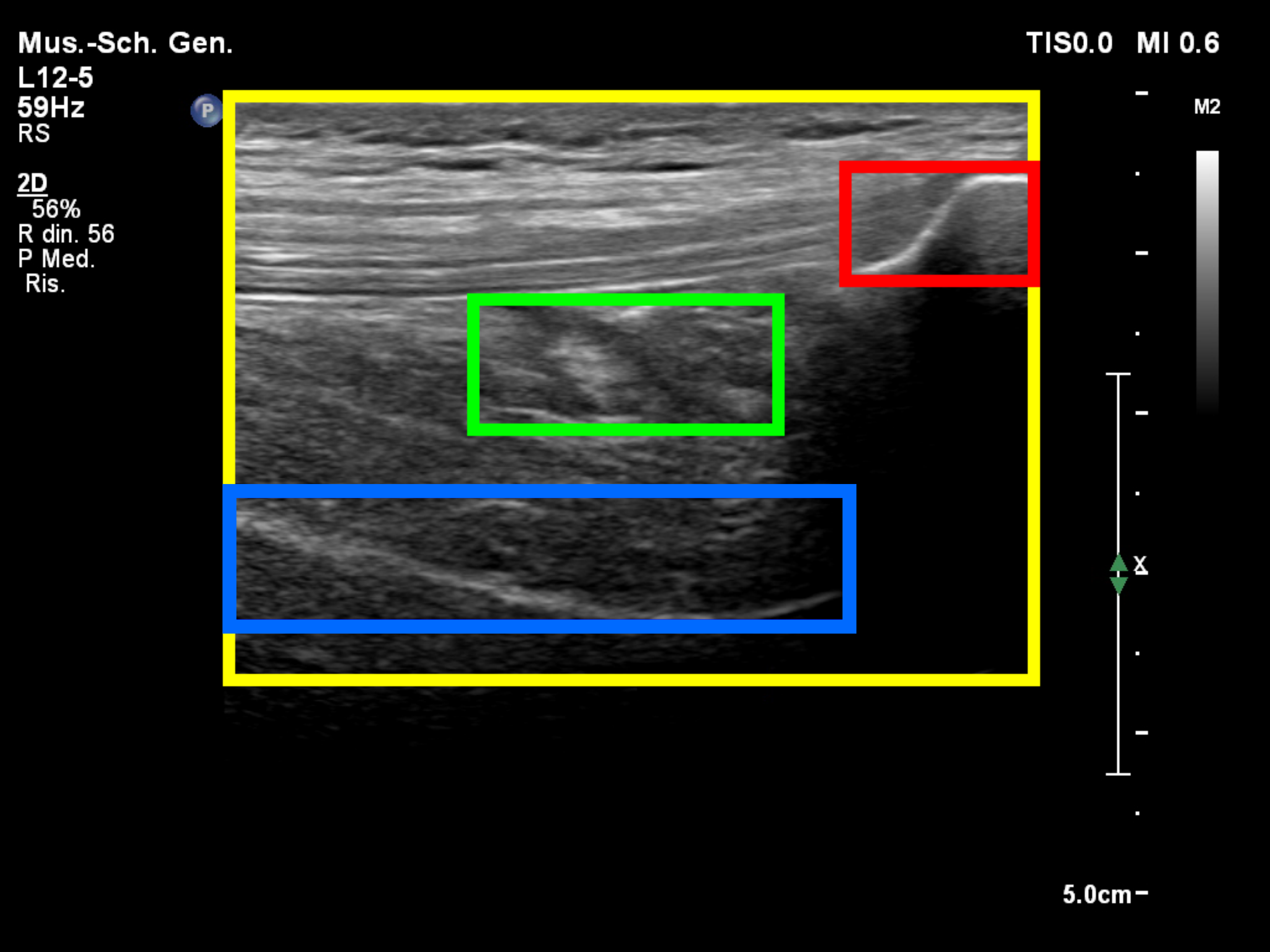}}
\hfill
\subfloat[Probe positioning\label{fig:pos}]{
\includegraphics[width=.4\textwidth, angle =90]{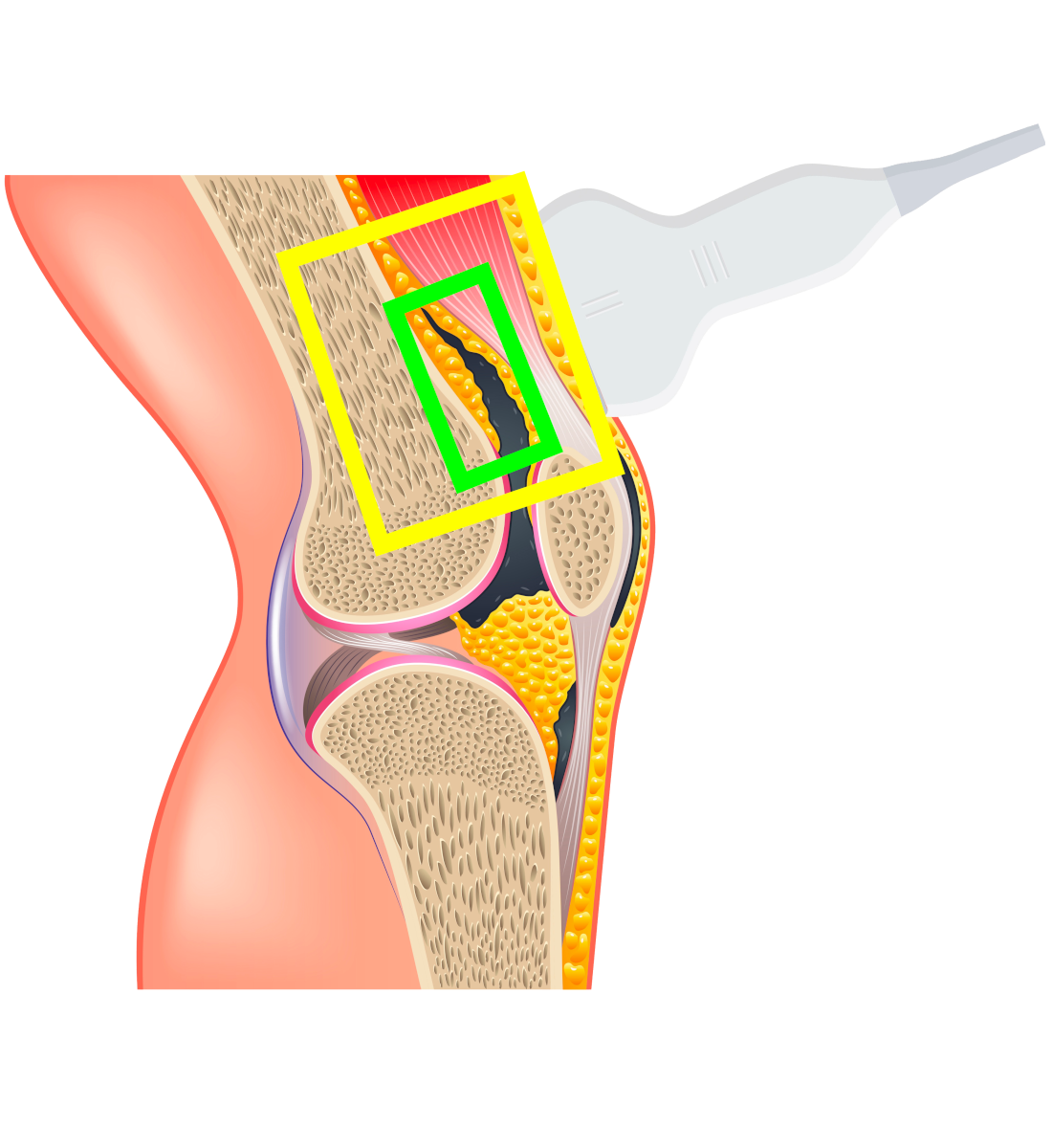}}
\caption{Image acquisition}
\label{fig:probe_pos}
\end{figure}

We focus on one of the three scans of the knee joint specified in HEAD-US protocol for the collection and diagnosis of joint recess distension in patients with hemophilia~\cite{MartinoliCarlo2013Dado}: the SQR longitudinal scan. 
This scan is used to assess SQR distension and contains different characterizing elements (see Figure \ref{fig:probe_pos}): 
\begin{itemize}
    \item The femur (blue box) usually appears as a light thick line, approximately horizontal, starting from the left side of the image and extending towards the right, often in the lower half of the image.
    \item The patella (red box) usually appears as a curved light line, positioned at the right border of the image, often in the top half and not entirely captured. The tendons (horizontal and parallel darker lines) can be seen on its left.
    \item The SQR (green box) often contains at least a small quantity of liquid and hence it is dark. In some cases, the joint recess membrane can be visible in grey. The joint recess is positioned between the femur and the patella. Its size and shape vary depending on many factors including whether it is distended or not, as explained below.
\end{itemize}

Figure~\ref{fig:pos} shows how the probe must be positioned during the acquisition of the SQR longitudinal scan. In the figure, the yellow box is the area that is captured by the US image shown in Figure~\ref{fig:example}, while the green box is the SQR. To correctly acquire this type of image, the knee has to be bent at 30$^\circ$. The probe must be positioned right at the beginning of the patella and moved horizontally to identify the correct key features previously described.

\subsection{Detection and classification of SQR distension}

\begin{figure}[ht] 
\centering
\subfloat[Non-distended SQR\label{fig:non_visible}]{
\includegraphics[width=.3\textwidth]{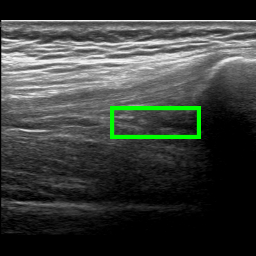}}
\hfill
\subfloat[Distended SQR\label{fig:visible}]{
\includegraphics[width=.3\textwidth]{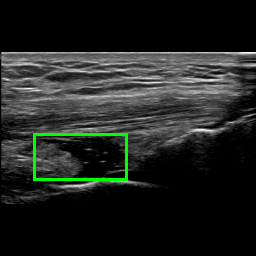}}
\hfill
\subfloat[Borderline Non-distended SQR\label{fig:borderline}]{
\includegraphics[width=.3\textwidth]{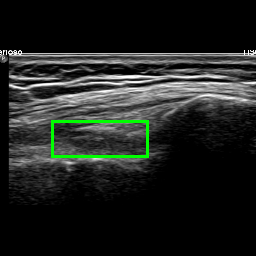}}
\caption{Examples of longitudinal SQR scans}
\label{fig:long_SQR}
\end{figure}

A joint recess can be distended due to three main reasons: it is filled with synovial liquid, it is filled with blood (a condition known as \textit{hemarthrosis}), and its membrane is thicker due to an inflammation known as \textit{synovitis}.
When the SQR is distended, it appears thicker on the US image. In some cases this can be clearly visible because the joint recess appears as a large dark area.
Figure~\ref{fig:long_SQR} shows three examples of the longitudinal SQR scan. In Figure~\ref{fig:non_visible} the SQR is the dark area shown in the green box. In this case the SQR is thin, hence it is not distended. Vice versa, in Figure~\ref{fig:visible} the SQR is much thicker, indicating that it is distended. While Figure~\ref{fig:non_visible} and \ref{fig:visible} show two characteristic examples with stark differences, there are borderline cases where the SQR appears slightly enlarged but it is not distended (see Figure~\ref{fig:borderline}) or it is very slightly distended.

An analysis, conducted with physicians from the Angelo Bianchi Bonomi Hemophilia and Thrombosis Center (two of which are also authors of this paper), revealed the need for a computer aided tool (CAD) supporting the physician in diagnosing SQR distension. The tool can be used as a part of a protocol for the early diagnosis of hemarthrosis, which is particularly relevant for hemophilic patients \cite{gualtierotti2021hemophilic, PlutDomen2019Daoh}.
Indeed, directly identifying hemarthrosis in US images is particularly challenging as it requires to distinguish blood from synovial fluid and blood clots from synovial hyperplasia, which appear very similar. 
%
To support the physician during the diagnosis, the CAD tool should identify the position of the SQR inside the specified US scan and classify it as \textit{Distended} or \textit{Non-distended}.
\section{Related work}

\subsection{US-based CAD systems}

Machine Learning (ML) techniques using medical imaging data have been investigated to support physicians in diagnosing various conditions~\cite{fujita2020ai}.
In particular, Ultrasound (US)~\cite{chan2011basics} is a very popular medical imaging technique, often used also as a data source for Computer-Aided Diagnosis (CAD) systems~\cite{huang2018machine, brattain2018machine}.
Indeed, despite its high dependence on the operator expertise level and possible noisiness of the acquired images~\cite{PlutDomen2019Daoh}, US imaging is easily accessible, safe and affordable and therefore commonly used in healthcare~\cite{brattain2018machine}.

In this problem domain, Convolutional Neural Networks (CNNs) are the most frequently used ML architectures, due to their ability to extract discriminative features from image data~\cite{ChenMin2021DFLf,simonyan2014very,akkus2019survey,sharma2018analysis}.
However, the development of such systems is often limited by the scarcity of available labeled data for the training of the ML models.
To mitigate this issue, in the literature, transfer learning~\cite{ChengPhillipM2017TLwC,TianjiaoLiu2017Cotn} and generative data augmentation approaches~\cite{Al-DhabyaniWalid2019DLAf,FujiokaTomoyuki2019BUIS} have been proposed.

\paragraph{\textbf{Classification approaches}}
One commonly used ML approach in US CAD systems is the direct classification of the images collected by medical experts~\cite{han2017deep, MengDan2017LFCB}. 
Indeed, different studies adopted deep learning classification approaches to identify various pathologies such as tumors in breast ultrasound~\cite{tanaka2019computer, becker2018classification, wang2020breast}, liver pathologies~\cite{acharya2015ultrasound, meng2017liver}, thyroid nodules~\cite{liu2017classification, song2019ultrasound}, and others~\cite{akkus2019survey}.

\paragraph{\textbf{Segmentation and detection approaches}}
Detection and segmentation techniques designed to extract Regions of Interest (ROI) in US images are also common.
For example, one solution  extracts a ROI of the femural carthilage from US images using segmentation~\cite{8857645}.
Other approaches rely on object detection architectures to detect multiple ROIs within a US image for example to
detect and classify breast lesions \cite{cao2019experimental} or to detect different types of diseases in several organs~\cite{zeng2020deep}.
Another example is SonoNet~\cite{baumgartner2017sononet}, a real-time detection network that identifies fetal standard scan planes in ultrasound 2D images.

\paragraph{\textbf{Multi-task learning approaches}}
Previous works have explored the multi-task combination of classification and detection for non-US medical images~\cite{yan2019mulan,gao2020feature,sainz2020multi,lin2017focal,le2019multitask}.
However, works exploring multi-task learning on US images are only a few.
Gong et al. propose an approach for multi-task localization of the thyroid gland and the detection of nodules within that region, using a shared backbone network which is divided into two different decoders for the two tasks~\cite{gong2021multi}.
Zhang et al. adopt a multi-task learning algorithm to segment and classify cancer in Breast US images. They propose to use DenseNet121 as backbone, followed by a decoder branch with layers connected by attention-gated (AG) units to segment the images~\cite{zhang2021sha}.
The second branch performs a classification task that takes in input the features extracted by the encoder.
We are not aware of multi-task learning algorithms adopted to address the problem of joint recess distension detection, and more generally we found no prior works proposing multi-task networks to analyze muskoskeletal US images.

\subsection{Joint recess distension detection approaches}
US images are commonly used for joint assessment and detection of joint recess distension in hemofilic patients. 
For this task, HEAD-US~\cite{MartinoliCarlo2013Dado} is a standardized protocol to support physicians in acquiring US images of commonly affected joints and formulating a diagnosis.

Two solutions have been proposed in the literature to automatically detect and classify joint recess distension but not focusing on patients with hemophilia, like in our contribution.
The former focuses on the shoulder joint~\cite{LinBor-Shing2020UDLi}, proposing the use of two CNNs to extract the ROI for Bicipital peritendinous effusions: the VGG-16~\cite{simonyan2014very} network is used for feature extraction, and a second CNN is used to classify the distension in three classes (i.e., mild, moderate, and severe).
The latter contribution considers the knee joints~\cite{long2020segmentation} and uses segmentation techniques to classify different pathologies inside US images, including joint recess distension due to synovial thickening.

A recent abstract paper~\cite{tyrrell2021detection} considers US images of patients with hemophila and  addresses the problem of classifying distended and not-distended knee recesses.
However, that prior work does not describe the methodology used for the classification, and it does not detect the joint recess ROI.

\section{Methodology} 

We propose two solutions for the problem defined in Section~\ref{sec:prob}.
The first solution, which we name \textit{Detection approach}, is described in Section~\ref{sub:detectionbased}.
It is based on a state-of-the-art detection technique, adopted to solve both the detection and the classification problems.
The second solution, which we call \textit{Multi-task approach} (see Section~\ref{sub:MTBoverview}), is a multi-task network with a branch that solves the detection problem and another one that solves the classification problem.

\subsection{\textit{Detection approach}}
\label{sub:detectionbased}
\begin{figure}[h!]
\centering
\includegraphics[width=.85\textwidth]{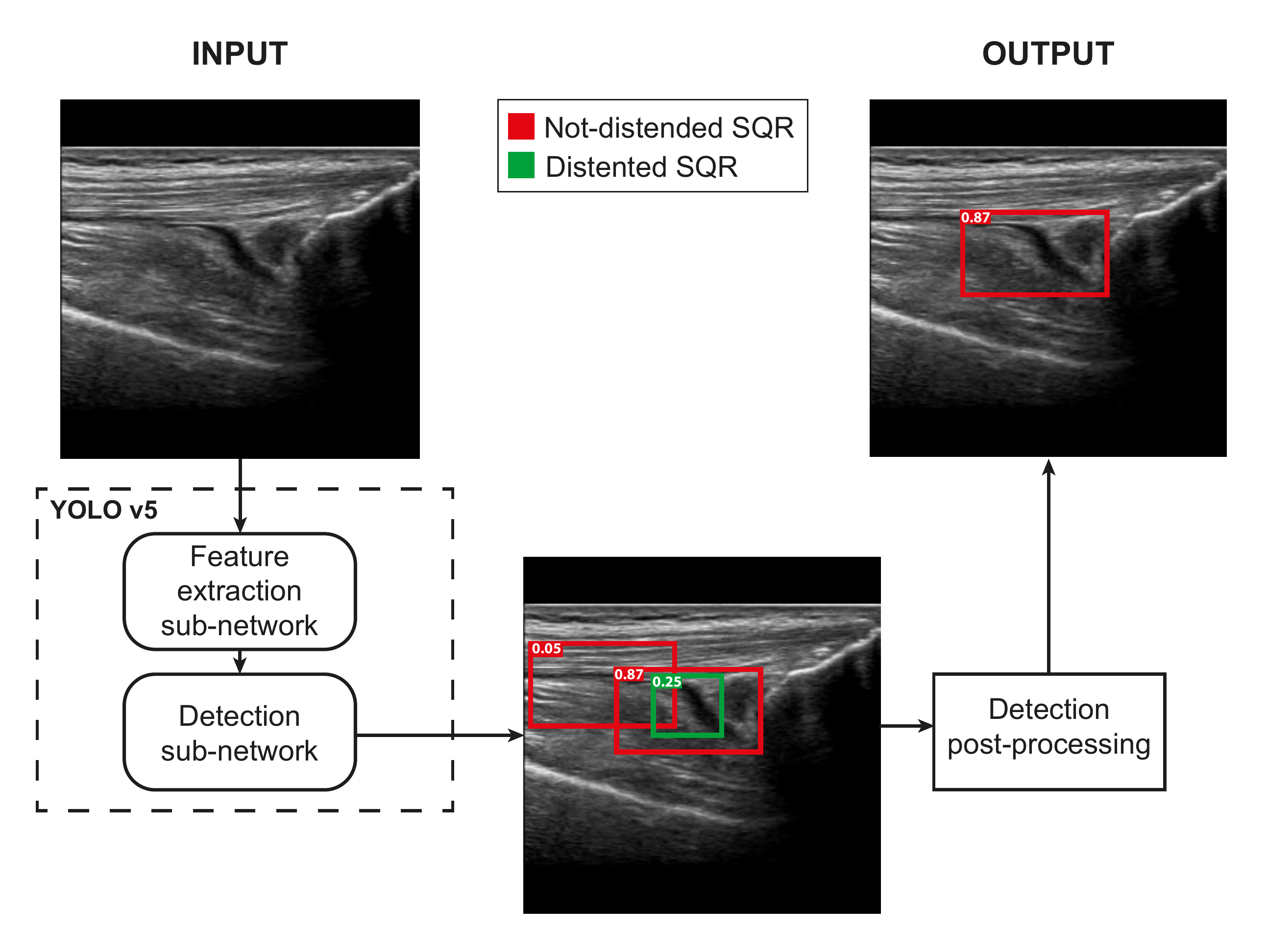} 
\caption{Overall architecture of the \textit{Detection approach}}
\label{fig:yolo_architecture}
\end{figure}

Figure~\ref{fig:yolo_architecture} depicts the network architecture of the \textit{Detection approach}. 
Each input US image is processed by the YoloV5~\cite{glenn_jocher_2022_6222936} object detector that returns a set of candidate SQRs, each characterized by a confidence value, a bounding box and the label (\textit{Distended} or \textit{Non-distended}).
Since in the considered domain the input image actually contains exactly one  SQR, the \textit{Detection Post-processing} module selects the prediction with highest confidence and outputs its bounding box and its label.

We train the network to recognize two classes of objects: \textit{Distended} SQRs and \textit{Non-distended} SQRs.
Since the amount of labeled images in
in this domain is generally scarce,
it is difficult to collect a sufficiently large dataset to fully train a robust detection network.
Therefore, we adopt a transfer learning approach~\cite{cheng2017transfer} to initialize the network's weights.
Specifically, we use the pre-trained weights publicly available
for the \textit{YoloV5} network trained on the MS COCO dataset~\cite{lin2014microsoft}. 
Finally, the network is fine-tuned on the actual dataset containing the labeled US images.

YoloV5 is a single stage detector designed to detect different objects in a image and directly assign them the corresponding class.
\textit{YoloV5} is an optimized version of the \textit{YoloV4} framework~\cite{bochkovskiy2020yolov4}, that has been widely used in the literature for object detection tasks.
Specifically, among the five models available in \textit{YoloV5}, we use the \textit{large} model, which was empirically selected.
%
\textit{YoloV5} is internally divided into a feature extraction sub-network and a detection sub-network. It also adopts a specific loss function and an early stop criterion. These four concepts are briefly described in the following.

\paragraph{\textbf{Feature Extraction sub-network}} The \textit{Feature Extraction sub-network} is a Convolutional Neural Network (CNN).
Specifically, it is a \textit{CSPDarknet53} network, that was originally proposed in~\cite{Chien-YaoWang2019CANB} and that was shown to be particularly effective for object detection~\cite{bochkovskiy2020yolov4} and ultrasound image classification~\cite{jabeen2022breast}.

\paragraph{\textbf{Detection sub-network}} The \textit{Detection sub-network} is divided into a \textit{neck} and a \textit{head} parts.
The overall goal of the \textit{neck} part is to divide the image into 
multiple small fragments
with the objective of simplifying further analysis by performing semantic segmentation (by associating categories to pixels) as well as instance segmentation (classifiying and locating objects at pixel level). 
The \textit{head} part is a one-stage detector~\cite{JosephRedmon2018YAII} that processes the features returned by the \textit{neck} part and outputs the bounding boxes of the detected elements along with their predicted class.

\paragraph{\textbf{Loss function}}
We use the default \textit{YOLOV5} loss function that is shown in Equation~\ref{eq:det_loss} and that is computed as the weighted sum of three values: a) the \textit{localization loss} ($L_{box}$) is computed with the \textit{Complete IoU} loss function (CIoU)~\cite{zheng2020distance}, and represents the error in the position of the predicted bounding box; b) the \textit{class loss} ($L_{c}$) is computed with Binary Cross-Entropy (BCE) and represents the error in classifying the predicted class; c) the \textit{objectness loss} ($L_{obj}$) is computed with BCE and represents to which extent the predicted bounding box actually encloses an object of interest.
The weights of these values are hyper-parameters that need to be empirically tuned (see Section~\ref{sub:hyper}). 

\begin{equation}\label{eq:det_loss}
L = \alpha  L_{box}+\beta L_{obj}+\gamma L_{c}
\end{equation}

\paragraph{\textbf{Early stopping criterion}}
We use the default \textit{YOLOV5} early stopping criterion to terminate the training if there are no improvements in the results for a given number of training epochs.
This default criterion considers the mean Average Precision (mAP) of the detection, \textit{i.e.}, the ratio of correctly classified bounding boxes considering a given threshold of the IoU with the corresponding ground truth.
Note that, in a multi-class scenario, this criterion factors for both the correct classification and the correct detection of the objects.
Specifically it is computed as the weighted sum of the mAP@0.5 and the mAP@0.5:0.95 where a weight of $0.1$ is given for mAP@0.5, and a weight of $0.9$ is given for mAP@0.5:0.95 in order to prioritize more accurate bounding boxes detection.

\subsection{\textit{Multi-task approach}}
\label{sub:MTBoverview}
The \textit{Detection approach} addresses the problem of classifying the SQR as distended or not by selecting the label of the detection with the highest confidence.
An alternative (and possibly more natural) solution would be to classify the entire image. However, this would not provide the needed SQR bounding box.
For this reason we propose the \textit{Multi-task Approach} that pairs image classification and detection.



\begin{figure}[h!]
\centering
\includegraphics[width=\textwidth]{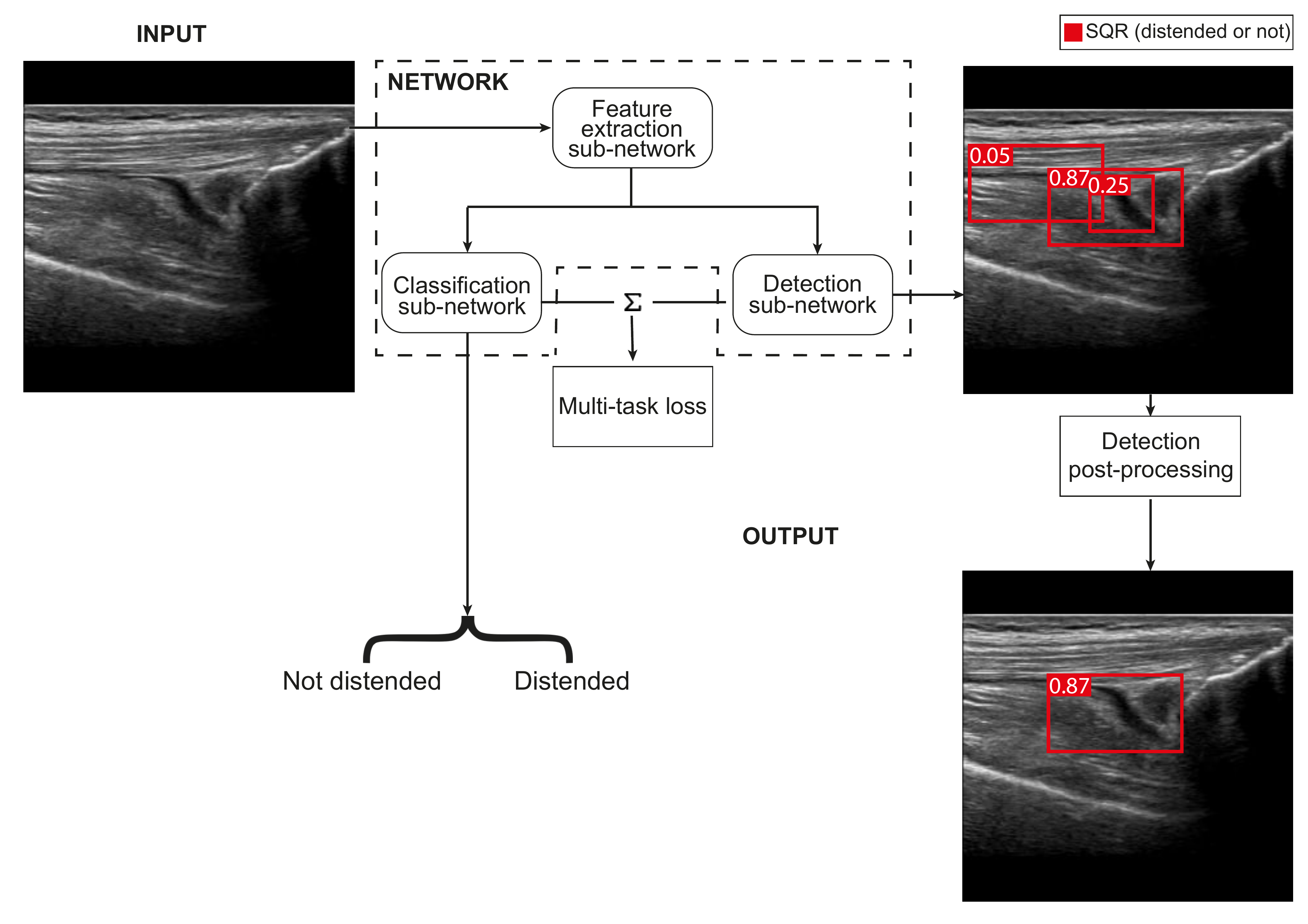} 
\caption{Overall architecture of the \textit{Multi-task approach}}
\label{fig:architecture}
\end{figure}

The proposed network is a modified version of the network used for the \textit{Detection approach}. The key modification consists of
a \textit{Classification sub-network} that performs the SQR binary classification.
The input image is first processed by the \textit{Feature Extraction} sub-network, that is shared for both classification and detection tasks. Then the extracted features are simultaneously processed by the \textit{Detection sub-network} and the \textit{Classification sub-network}. The \textit{Classification sub-network} processes the features and returns the predicted SQR class (i.e., distended or not) considering the whole image.

Differently from the \textit{Detection Approach} solution, the goal of the \textit{Detection sub-network} in the \textit{Multi-task} solution is simply to detect the SQR, without providing information about the distension.
Hence, the \textit{Detection sub-network} network is trained with a single class and it returns a set of bounding boxes, all belonging to the same class, each with an associated confidence value. The \textit{Detection Post-processing} module selects the bounding box with the highest confidence.
During the training phase, 
the \textit{multi-task loss} jointly considers the errors on classification and detection to update the network weights.

\subsubsection{Classification sub-network}
\label{sub:MTBclassification}

\begin{figure}[h!]
\centering
\includegraphics[width=.75\textwidth]{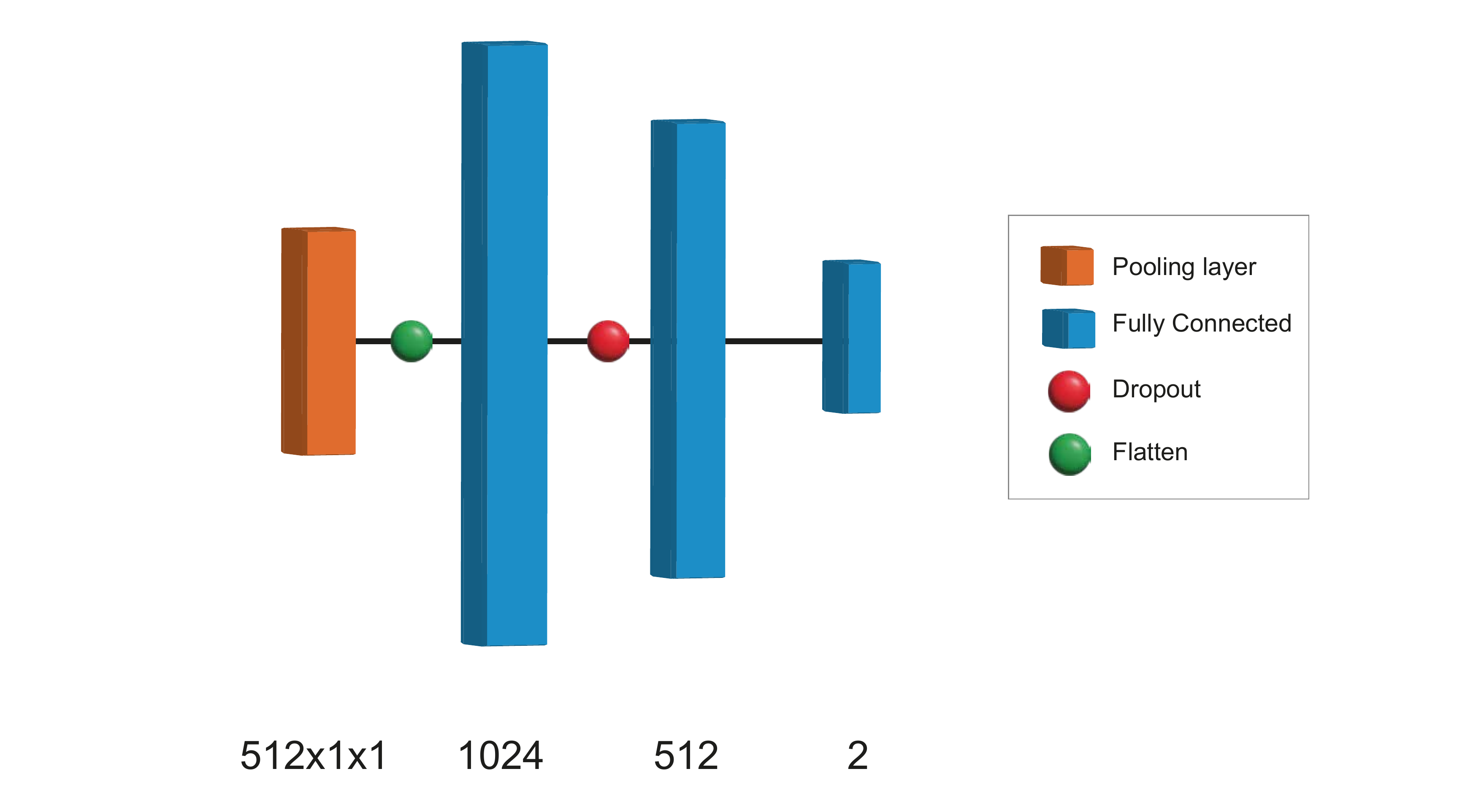} 
\caption{Classification sub-network architecture}
\label{fig:cls_architecture}
\end{figure}

Figure~\ref{fig:cls_architecture} shows the \textit{Classification sub-network} of the \textit{Multi-task Approach}.
The first layer of the sub-network is an Adaptive Average Pooling Layer in charge of reducing the feature dimensions to a fixed 2-dimensional output size.
Then, the output is provided to a Flatten Layer, that converts 2-dimensional data to a 1-dimensional array.
This array is then processed by a fully connected network composed of two hidden layers of $1024$ and $512$ units, respectively. These layers use a \textit{ReLu} activation function. A dropout layer is applied between the two hidden layers with the objective of reducing overfitting. Finally, a Softmax layer is in charge of providing the most likely class (i.e., \textit{Distended}/\textit{Non-distended}). The architecture of this network has been determined empirically.

\subsubsection{Multi-task loss}
\label{sub:loss}
Training the multi-task network requires a custom loss function that simultaneously takes into account the classification and detection errors. 
For this reason, we adapt the loss function used for the \textit{Detection approach} by adding a new loss term that represents the errors of the \textit{Classification sub-network}.
Specifically, we adopt a typical solution in binary classification that consists in computing the classification error $L_{cls}$ with a BCE function.
Another difference with respect to the loss function used in the \textit{Detection approach}, is that, in the \textit{Multi-Task Approach},  the \textit{Detection sub-network} is trained with a single class, hence there are no possible errors with class prediction. Thus, the $L_{c}$ parameter, considered in Equation~\ref{eq:det_loss}, is always zero.
So, the overall multi-task loss is computed as the weighted sum of $L_{box}$, $L_{obj}$, and $L_{cls}$, as shown in Equation~\ref{eq:mtl_loss}.
These weights are hyper-parameters that need to be empirically tuned (see Section~\ref{sub:hyper}). 



\begin{equation}\label{eq:mtl_loss}
L = \alpha  L_{box}+\beta L_{obj}+\delta L_{cls} 
\end{equation}


Since the datasets in this domain are usually highly unbalanced (e.g., in our dataset $\approx75\%$ of the images are labeled as \textit{Non-distended}),
there is the risk that the network favors \textit{Non-distended} classifications, which in turn may increase the number of false negatives. 
In order to mitigate this problem, we adjust the classification loss $L_{cls}$ to give higher error values to false negatives (i.e., \textit{Distended} SQR classified as \textit{Non-distended}).
This is achieved by adding an additional weight to $L_{cls}$ when the ground truth is \textit{Distended}. In this case the weight is the ratio between the \textit{Non-distended} and \textit{Distended} samples in the training set.
Thanks to this approach, the errors on the \textit{Distended} samples have a more significant impact on the overall loss.

\subsubsection{Multi-task early stopping criterion}
\label{sub:earlystopping}

As specified above, for the \textit{Detection approach}, the default \textit{YOLOV5} early stopping criterion, based on mAP, is used to stop the training if no improvements are detected for a specified number of epochs.
Instead, for the \textit{Multi-task approach},
since the detection is computed for a single class, the mAP does not account for the classification accuracy but only considers the detection accuracy.
Thus, for the \textit{Multi-task approach}, we consider a weighted sum of mAP@0.5 for the detection and Balanced Accuracy (BA) for the classification on the validation set. In particular, we provide a higher weight ($0.7$) to the Balanced Accuracy and a lower one to to mAP@0.5 ($0.3$).
This is due to the fact that we prefer to be more accurate on the classification, at the cost of identifying slightly less accurate (but still informative) bounding boxes. We consider a patience value of $100$ epochs, which means that the training is stopped if the early stopping criterion does not improve for the number of epoch specified by the patience value.
\section{Dataset}\label{sec:dataset}
Due to the lack of publicly available datasets in this field, we collected a dataset of SQR longitudinal scan images of $200$ adult patients with hemophilia, aged $44.7\pm18.6$, between January 2021 and May 2022, thanks to the collaboration with 'Centro Emofilia e Trombosi Angelo Bianchi Bonomi' of the polyclinic of Milan, a medical institution specialized in hemophilia.
The study was approved by the institution's ethics committee. 

Before acquiring the dataset we first defined a standardized data acquisition protocol that includes: a) examination procedure based on the HEAD-US~\cite{MartinoliCarlo2013Dado} protocol; b) guidelines on how to use the ultrasound device during the visit, for example defining that the joint side (left or right) should be annotated while acquiring the image itself; c) a procedure for data extraction from the ultrasound device; d) a data pseudo-anonymization procedure. 

For each patient, the physician collected several US images from various scans in different joints. For this study we selected images of the SQR longitudinal scan. Two images of the SQR longitudinal scan are typically collected during each visit, one for each knee (left/right) but for some patients we only have one image while other patients were visited twice (often at a distance of several months), hence having up to four images each.

\subsection{Data Acquisition}

Images were acquired using the Philips Affiniti 50 US device\footnote{\url{www.usa.philips.com/healthcare/product/HC795208/affiniti-50-ultrasound-system}} by a single specialized practitioner during routine visits of hemophilic patients.
When collecting the images, the probe was positioned as shown in Figure~\ref{fig:pos} and the knee was flexed by $30^{\circ}$. 
Each image has a resolution of $1024\times 780$ and, as shown in Figure~\ref{fig:example}, it contains acquisition parameters (saved as text in the image) and the actual US scan (\textit{i.e.}, the yellow rectangle in Figure~\ref{fig:example}). Note that the position and the dimension of the actual US scan (\textit{i.e.}, the yellow box) can vary.
During the annotation process, the practitioner reported for each image whether the SQR was distended or not.
Additionally, a trained operator, supervised by the practitioner, annotated the Minimum Bounding Rectangle (MBR) of the SQR.

A total of 483 knee subquadricipital recess longitudinal scans were annotated, 360 of which are labeled as \textit{Non-distended} and 123 as \textit{Distended}.

\subsection{Pre-processing}\label{sec:cropping}
We pre-process the collected images to extract the actual US image (e.g., the yellow box in Figure~\ref{fig:example}). 
Indeed, as previously observed \cite{LinBor-Shing2020UDLi,long2020segmentation} using the entire image as returned by the US device can reduce classification accuracy as this part of the image does not contain information needed for the required tasks.

\begin{figure}[ht] 
 \centering
 \includegraphics[width=0.85\textwidth]{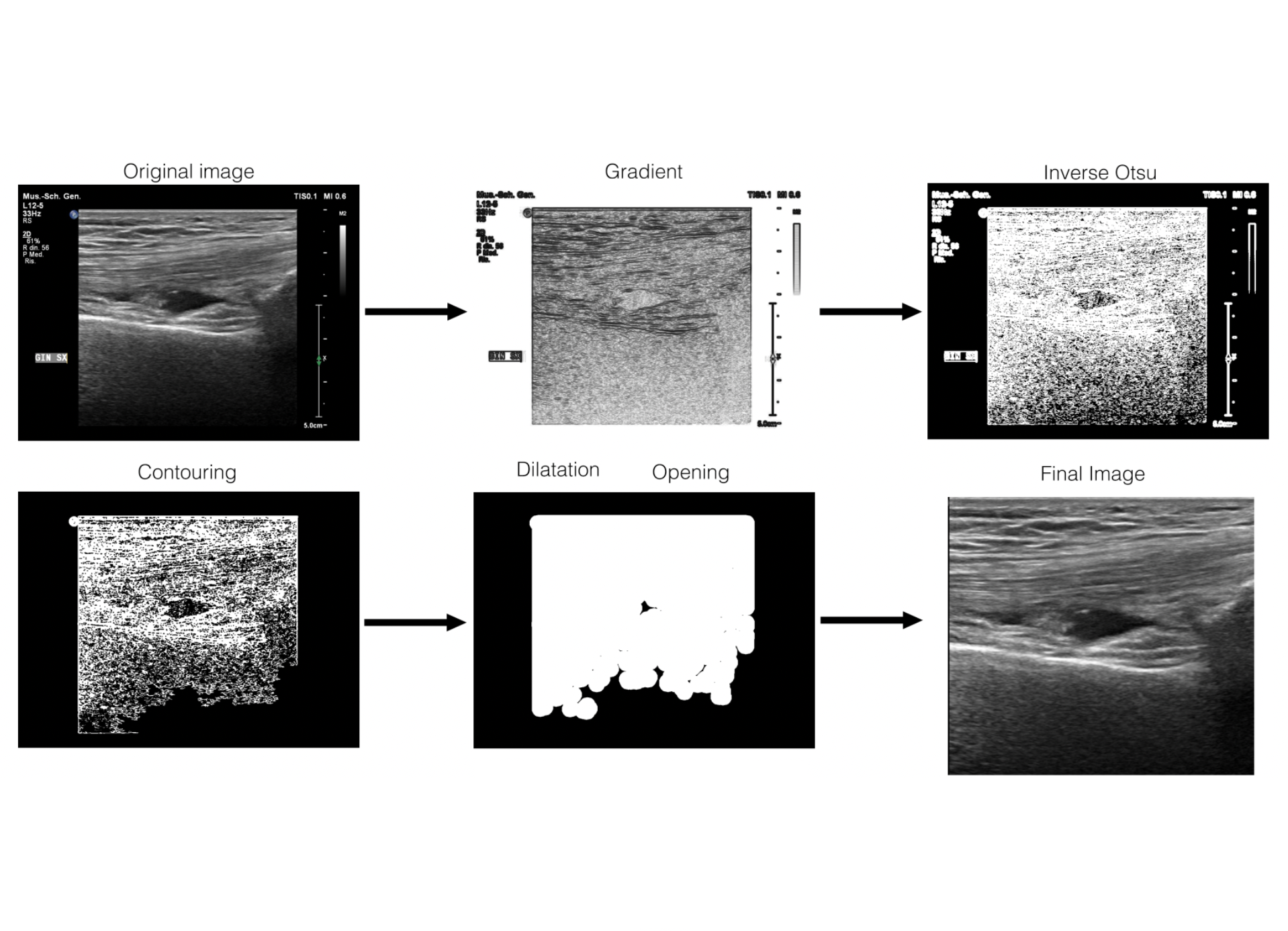}
 \caption{Intermediate steps of frame extraction procedure}
\label{fig:crop_proc}%
\end{figure}

As suggested by Tingelhoff et al~\cite{tingelhoff2008analysis} we initially cropped the images manually.
However, this process is time consuming. 
We therefore developed an algorithm to automatically extract the US scan from the collected image.
Figure~\ref{fig:crop_proc} shows the steps of the pre-processing algorithm.
In the first step, we measure and binarize the gradient of the image; we then remove connected pixel groups composed of less than $1000$ non-zero pixels; afterwards, we dilate the image to fill small groups of black pixels, and we perform an opening operation to remove groups of pixels not belonging to the US scan that were merged with it in the previous steps. Finally we crop the original image with the bounding box of the white area resulting from the previous step.
Finally the images are resized to $256\times256$ pixels.

All images have been double-checked as part of the annotation process and no cropping error was found, showing that the proposed automatic pre-processing is reliable. 

\section{Evaluation}

In this section, we describe our experimental evaluation on the dataset that we introduced above. First, we present the adopted evaluation methodology and the metrics. Then, we describe how we selected the hyper-parameters. Finally, we show and compare the results of the two proposed solutions and present some examples.

\subsection{Metrics}
We define two sets of metrics: one for the detection and the other for the classification.
For what concerns the detection, we measure the average Intersection over Union (IoU). The IoU between two plane figures is defined as the ratio between the area of their intersection and the area of their union. For each test image we measure the IoU between the bounding box predicted with the highest confidence and the ground truth bounding box. Then, we compute the average of this metric among all images.

Considering classification, for each image we compare the ground truth class with the predicted class hence computing if the result is a True Positive (TP), a True Negative (TN), a False Positive (FP),  or a False Negative (FN). Note that the positive class is \textit{Distended} and the negative class is \textit{Non-distended}. Then, we used the following classification metrics:




\begin{itemize}
    \item Specificity: measures the ability of the model to identify true negatives. Specificity$={TN\over {TN+FP}}$
    \item Sensitivity: measures the ability of the model to identify true positives.
    Sensitivity$={TP\over {TP+FN}}$
    \item Balanced Accuracy: mean between specificity and sensitivity. It is considered a sounder metric compared to accuracy when the class imbalance is high~\cite{brodersen2010balanced}. Balanced accuracy $={{sens+spec}\over2}$
\end{itemize}

\subsection{Evaluation methodology}\label{sec:eval_m}
The evaluation of the recognition rate of the proposed solutions is based on a 5-fold cross-validation.
In order to avoid high correlation bias, the training and the test splits do not have images from the same patients in common.
The consequence is that we could not exactly divide the dataset in $80\%$ and $20\%$ splits and therefore the splits have slightly different number of images.
%
An example fold subdivision can be found in Table~\ref{tab:fold_0}.
Each training fold was further split: $80\%$ as training set and $20\%$ as validation set. During training we used SGD with momentum~\cite{sutskever2013importance} as optimizer.


\begin{table}[h!]
\centering
\begin{tabular}{|l|r|r|r|}
\hline
\textbf{Fold 0} & \multicolumn{1}{l|}{Train} & \multicolumn{1}{l|}{Test} & \multicolumn{1}{l|}{Total} \\ \hline
Non-distended   & 289 & 71 & 360 \\ \hline
Distended & 97 & 26 & 123 \\ \hline
Total & 386 & 97 & 483 \\ \hline\hline
Total patients & 166 & 42 & 208 \\ \hline
\end{tabular}
\caption{Example data distribution in Fold 0 of 5-f CV}
\label{tab:fold_0}
\end{table}

\subsection{Hyper-parameters selection}
\label{sub:hyper}
In order to properly tune the many hyper-parameters of our network, we adopt an evolutionary approach~\cite{bochinski2017hyper}.
Given a fitness function, an evolutionary algorithm evaluates the best fitting set of hyper-parameters thanks to \textit{mutation} and \textit{cross-over} operations.
For the sake of this work, we considered the evolutionary method proposed in \textit{YOLOV5}, that only considers the mutation operation with $90\%$ of probability and $0.04$ of variance. Each mutation step generates a new set of hyper-parameters given a combination of the best parents from all the previous generations. The fitness functions used for the hyper-parameters selection for the \textit{Detection approach} and the \textit{Multi-task approach} correspond to the early stopping criteria introduced in Sections~\ref{sub:detectionbased} and \ref{sub:earlystopping}, respectively. 

In order to balance the need for a high number of evolution epochs with limited computational resources, we run the evolutionary algorithm only on one fold. 

We executed our evolutionary algorithm for $300$ epochs on each solution.
Considering the \textit{Multi-task Approach}, the best results have been obtained at the $193$th epoch, while for the \textit{Detection approach} the best set of hyper-parameters were found at the $4$th epoch.
The set of hyper-parameters resulting from evolution have been used to evaluate our approaches on the complete cross validation procedure. The most relevant discovered hyper-parameters are presented in Table~\ref{tab:hyper}

\begin{table}[h!]
\resizebox{\textwidth}{!}{
\begin{tabular}{|c|c|c|c|c|c|c|c|}
\hline & Learning rate  & Dropout & SGD momentum & $\alpha$ & $\beta$ & $\gamma$ & $\delta$ \\ \hline
Detection  & 0.00369 & - & 0.77628  & 0.06868 & 0.49062 & 0.2343  & - \\ \hline
Multi-task & 0.0018 & 0.11008 & 0.62403  & 0.05427 & 0.67598 & - & 0.41855 \\ \hline
\end{tabular}}
\caption{Selected hyper-parameters}
\label{tab:hyper}
\end{table}

Note that $\gamma$ is a weight associated to the $L_{c}$ loss that is only considered in the \textit{Detection approach}, while $\delta$ is a weight associated to the $L_{cls}$ loss that is only considered in \textit{Multi-task approach}. Finally, the Dropout rate is only included in the \textit{Classification sub-network} of the \textit{Multi-task approach}.




\subsection{Results}
Table~\ref{tab:res} shows the performance of the two proposed solutions.
\begin{table}[h!]
\resizebox{\textwidth}{!}{
\begin{tabular}{|l|l|l|l|l|} \hline
 & Balanced accuracy & Specificity & Sensitivity & IoU\\ \hline
\textit{Detection Approach} & 0.74 $\pm$ 0.07 & \textbf{0.97 $\pm$0.03} & 0.52 $\pm$ 0.12 & \textbf{0.66 $\pm$ 0.01} \\ \hline
\textit{Multi-task Approach} & \textbf{0.78 $\pm$ 0.05} & 0.92 $\pm$ 0.04 & \textbf{0.64 $\pm$ 0.09} & 0.63 $\pm$ 0.02 \\ \hline
\end{tabular}}
\caption{Evaluation results (reported as mean among the folds $\pm$ standard deviation)}
\label{tab:res}
\end{table}
Since both the early stopping criterion and the hyper-parameters selection methods for the \textit{Multi-task approach} are designed to prioritize the classification accuracy at the expense of the detection accuracy, its balanced accuracy is confirmed to be more accurate with respect to the \textit{Detection approach}.
This increase is particularly relevant as it brings the balanced accuracy over the threshold of $0.75$ that is reported to be a requirement for a medical test to be ``useful''~\cite{power2013principles}.

The increase in the balanced accuracy comes at the expense of a reduction in the detection accuracy.
Indeed, the IoU metric decreases from $0.66$ for the \textit{Detection Approach} to $0.63$ for the \textit{Multi-task Approach}.
However, if we consider the percentage of the images for which the IoU between the detection and the ground truth is $\geq 0.5$, which is a common threshold used for similar tasks, we notice that the decrease is relatively small.
Indeed, the result for the \textit{Detection Approach} is $85\%$ and it decreases to $82\%$ for the \textit{Multi-task Approach}.

\begin{figure}[h!]
\centering
\subfloat[Multi-task learning\label{fig:cm_mtl}]{
\includegraphics[width=.48\textwidth]{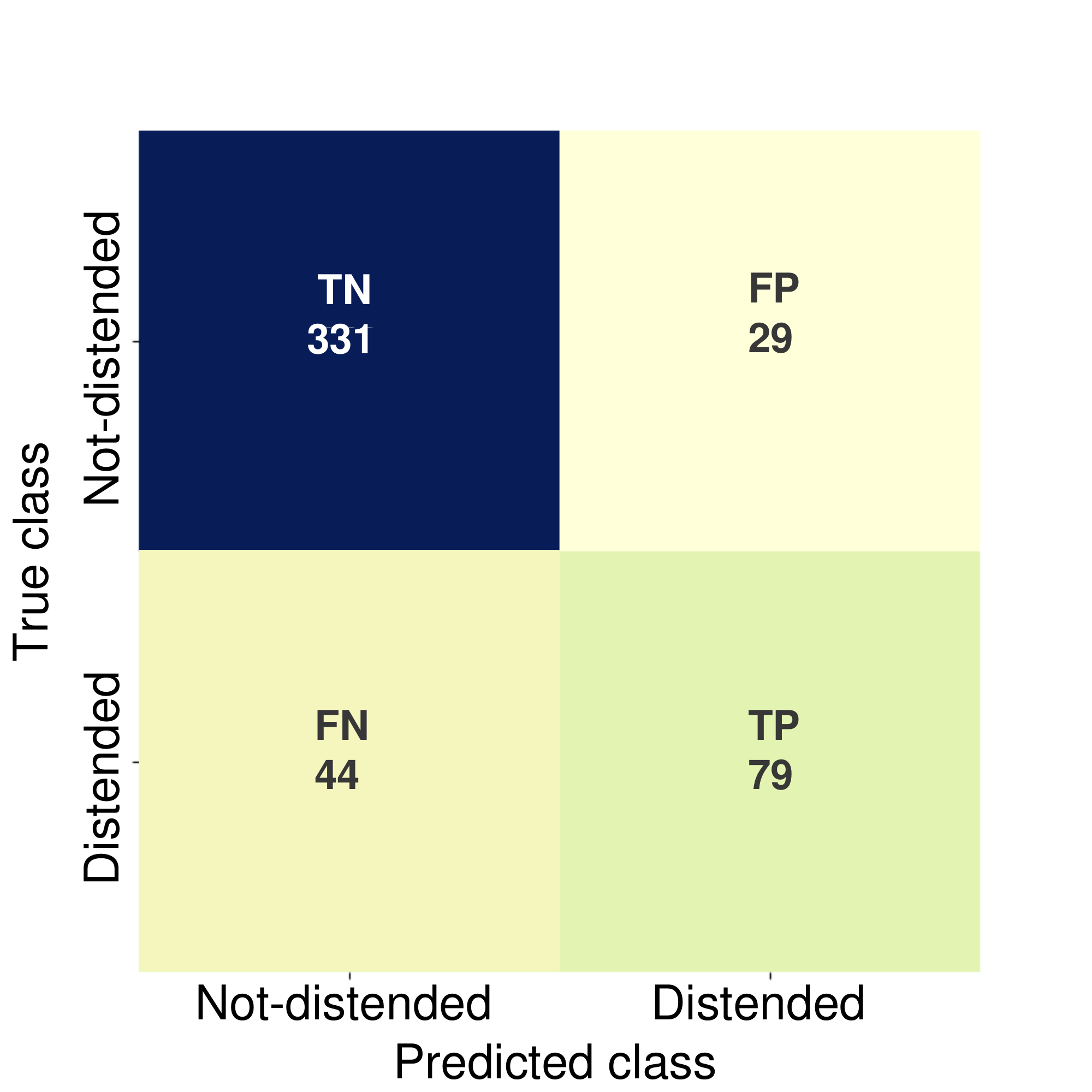}}
\hfill
\subfloat[Detection\label{fig:cm_yolo}]{
\includegraphics[width=.48\textwidth]{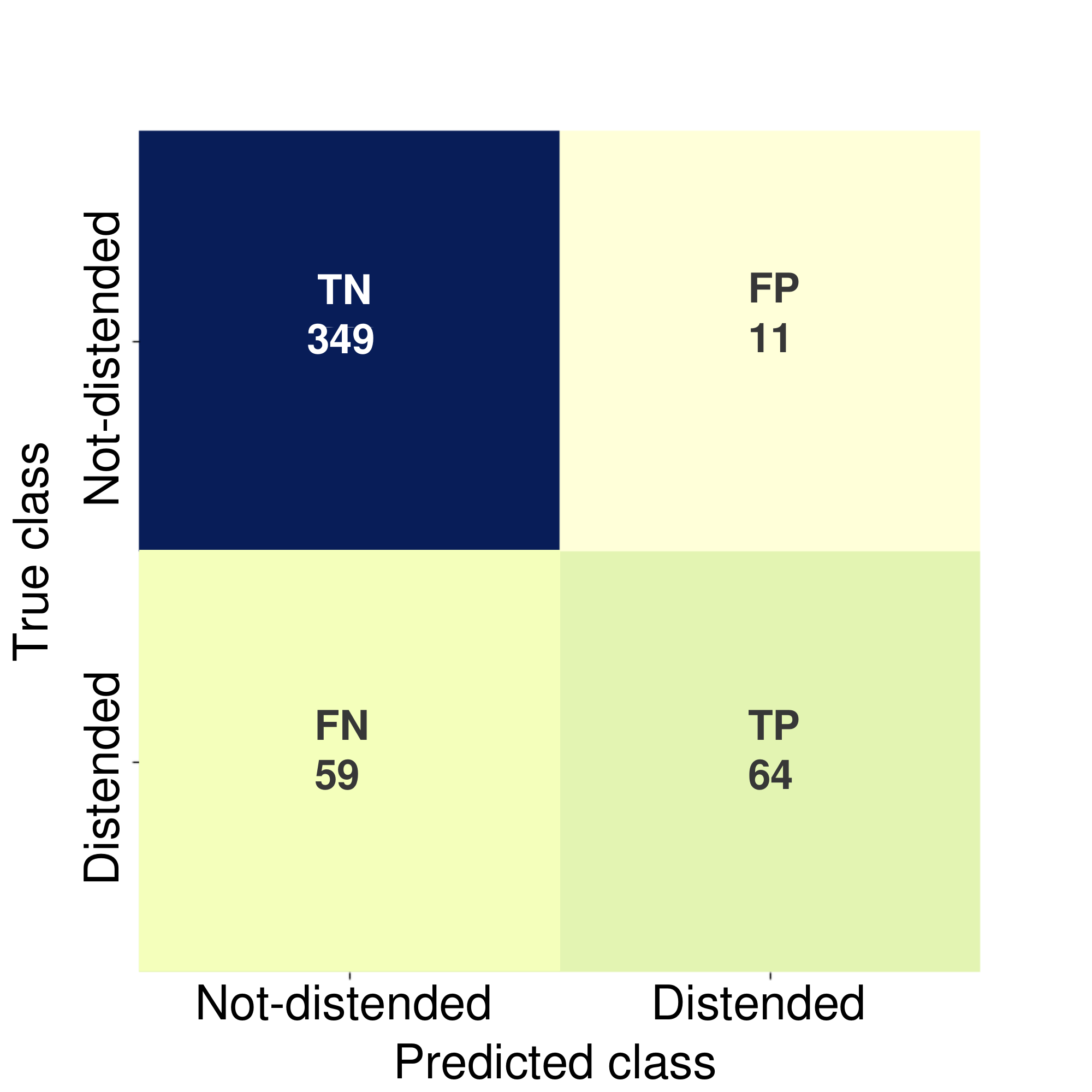}}
\caption{Confusion matrices}
\label{fig:confusion_matrix}
\end{figure}

The increase in balanced accuracy value of the \textit{Multi-task Approach} is largely influenced by the increase in \textit{sensitivity}.
The reason for this increase is likely due to the adjusted classification loss in the \textit{Multi-task Approach} introduced to mitigate the unbalanced data problem (see Section~\ref{sub:loss}).
Indeed, considering the confusion matrices in Figure~\ref{fig:confusion_matrix} we can observe that the \textit{Detection Approach} has $59$ false negatives ($48\%$), out of a total of $123$ images labelled as \textit{Distended}, compared to the $44$ false negatives in the \textit{Multi-task Approach} ($38\%$).
This improvement comes at a cost of a lower \textit{specificity} value.
Indeed, the \textit{Detection Approach} has $11$ false positives out of $360$ negative images ($3\%$) while the \textit{Multi-task Approach} has $29$ false positives ($8\%$).

\subsection{Examples}
In order to better illustrate how our approaches work, in the following we show some examples of correct and incorrect output.

\begin{figure}[h!] 
\centering
\subfloat[\textit{Non-Distended} SQR\label{fig:hit_0_0}]{
\includegraphics[width=.45\textwidth]{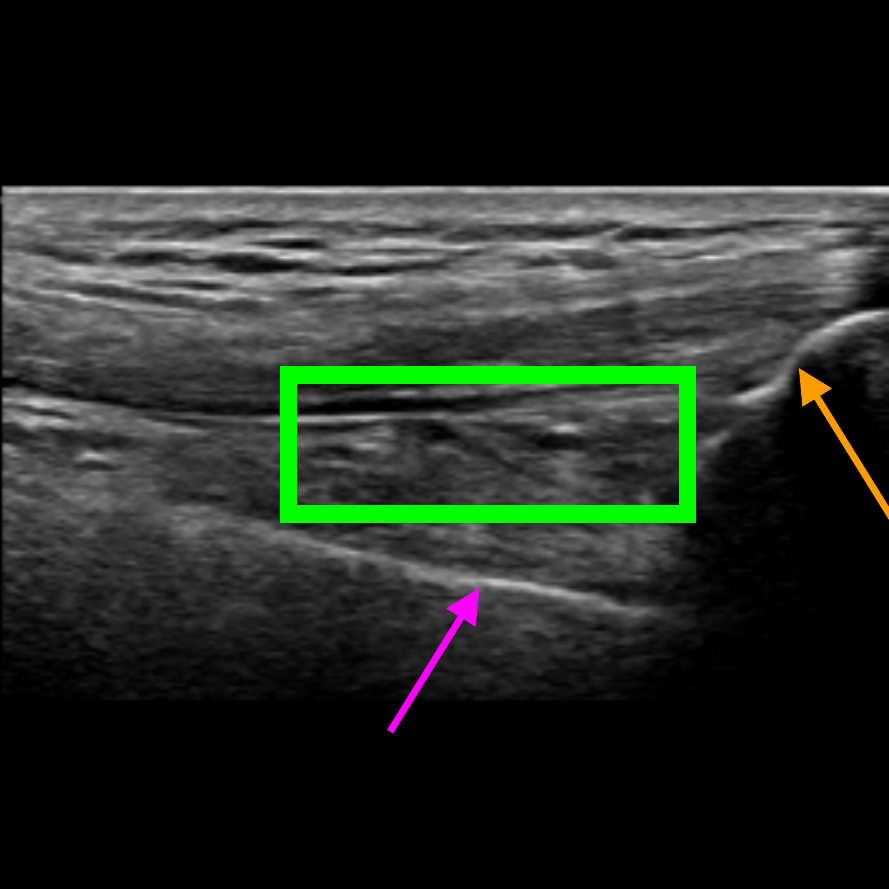}}
\hfill
\subfloat[\textit{Distended} SQR\label{fig:hit_1_1}]{
\includegraphics[width=.45\textwidth]{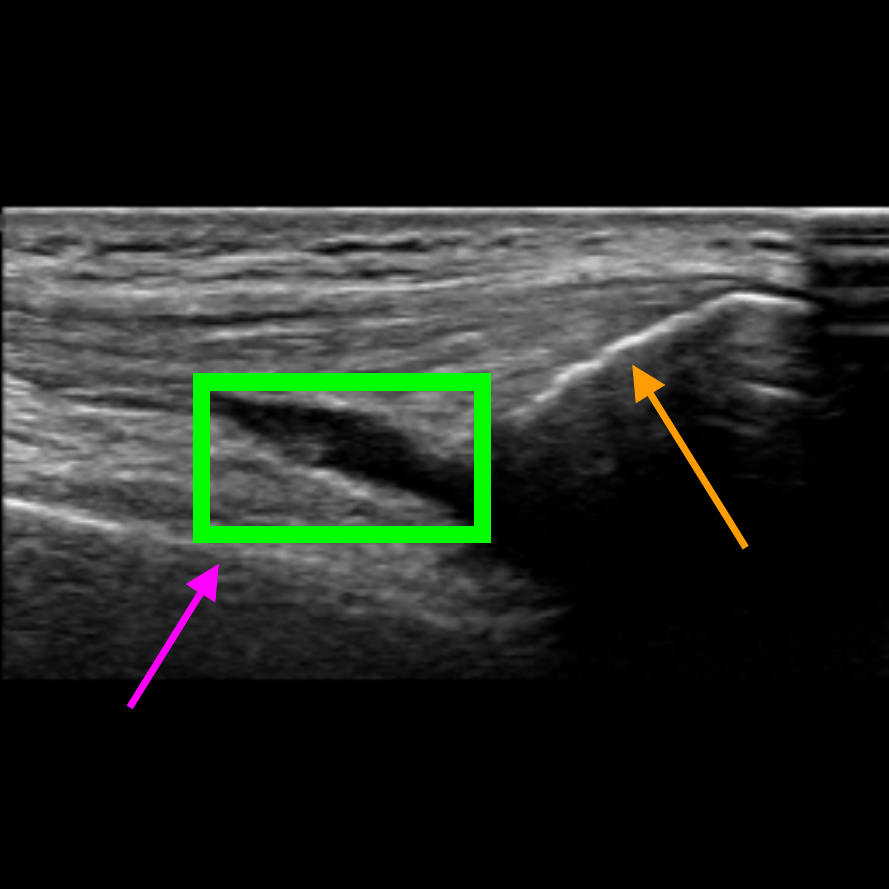}}
\caption{Examples of images correctly classified by both solutions. The purple arrow points to the femur, the orange arrow points to the patella, and the green box indicates the SQR.}
\label{fig:class_correct}
\end{figure}

Figure~\ref{fig:class_correct} shows two US images that have been correctly classified by both approaches and that are relatively easy to classify by medical experts.
Figure~\ref{fig:hit_0_0} shows an 
US image where
the femur, the patella and the SQR are clearly visible, and the SQR is thin (\textit{i.e.}, not distended).
On the other hand, Figure~\ref{fig:hit_1_1} shows an  example of a \textit{Distended} SQR. In this case, the SQR is clearly thick and hence distended.

\begin{figure}[h!] 
\centering
\subfloat[\textit{Non-distended} SQR\label{fig:hard_cc}]{
\includegraphics[width=.4\textwidth]{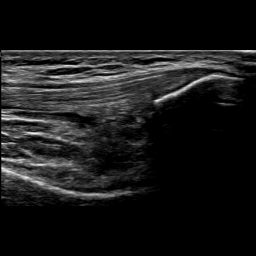}}
\hfill
\subfloat[\textit{Distended} SQR\label{fig:hard_yw}]{
\includegraphics[width=.4\textwidth]{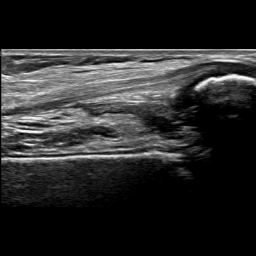}}
\hfill
\subfloat[\textit{Non-distended} SQR\label{fig:hard_mw}]{
\includegraphics[width=.4\textwidth]{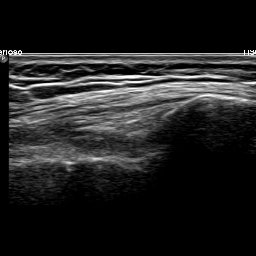}}
\hfill
\subfloat[\textit{Distended} SQR\label{fig:hard_ww}]{
\includegraphics[width=.4\textwidth]{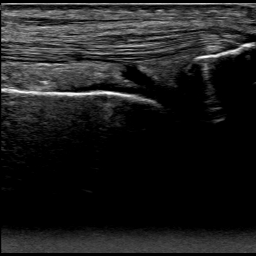}}
\caption{Examples of images that are intuitively hard to classify.}
\label{fig:hard}
\end{figure}

Figure~\ref{fig:hard} shows four examples of images that are more challenging to classify even by medical experts. This usually happens when there is noise in the US scan (as in Figure~\ref{fig:hard_mw}) or when the SQR is borderline between \textit{Distended} and \textit{Non-distended} (as in Figure~\ref{fig:hard_ww}).
Figure~\ref{fig:hard_cc} is correctly classified by both approaches as \textit{Non-distended}. Figure~\ref{fig:hard_yw} is correctly classified by the \textit{Multi-task approach} but not by the \textit{Detection approach}.
Vice versa, Figure~\ref{fig:hard_mw} is correctly classified by the \textit{Detection approach} and not by the \textit{Multi-task approach}. Finally, both solutions wrongly classify Figure~\ref{fig:hard_ww}.

\begin{figure}[h!]
\centering
\subfloat[Worst detection by \textit{Multi-Task approach}\label{fig:min_iou_multi}]{
\includegraphics[width=.4\textwidth]{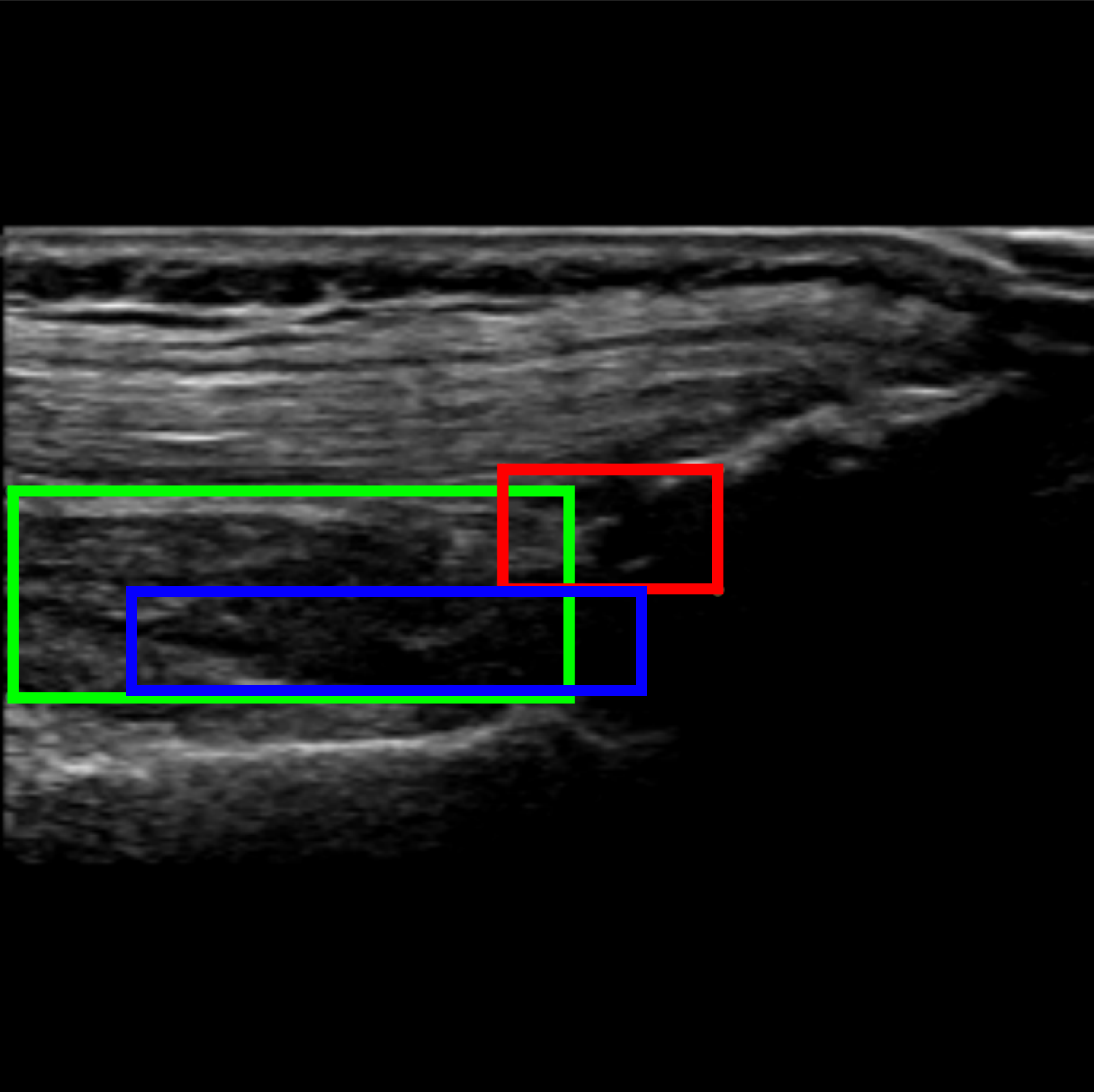}}
\hfill
\subfloat[Best detection by \textit{Multi-Task approach}  \label{fig:max_iou_multi}]{
\includegraphics[width=.4\textwidth]{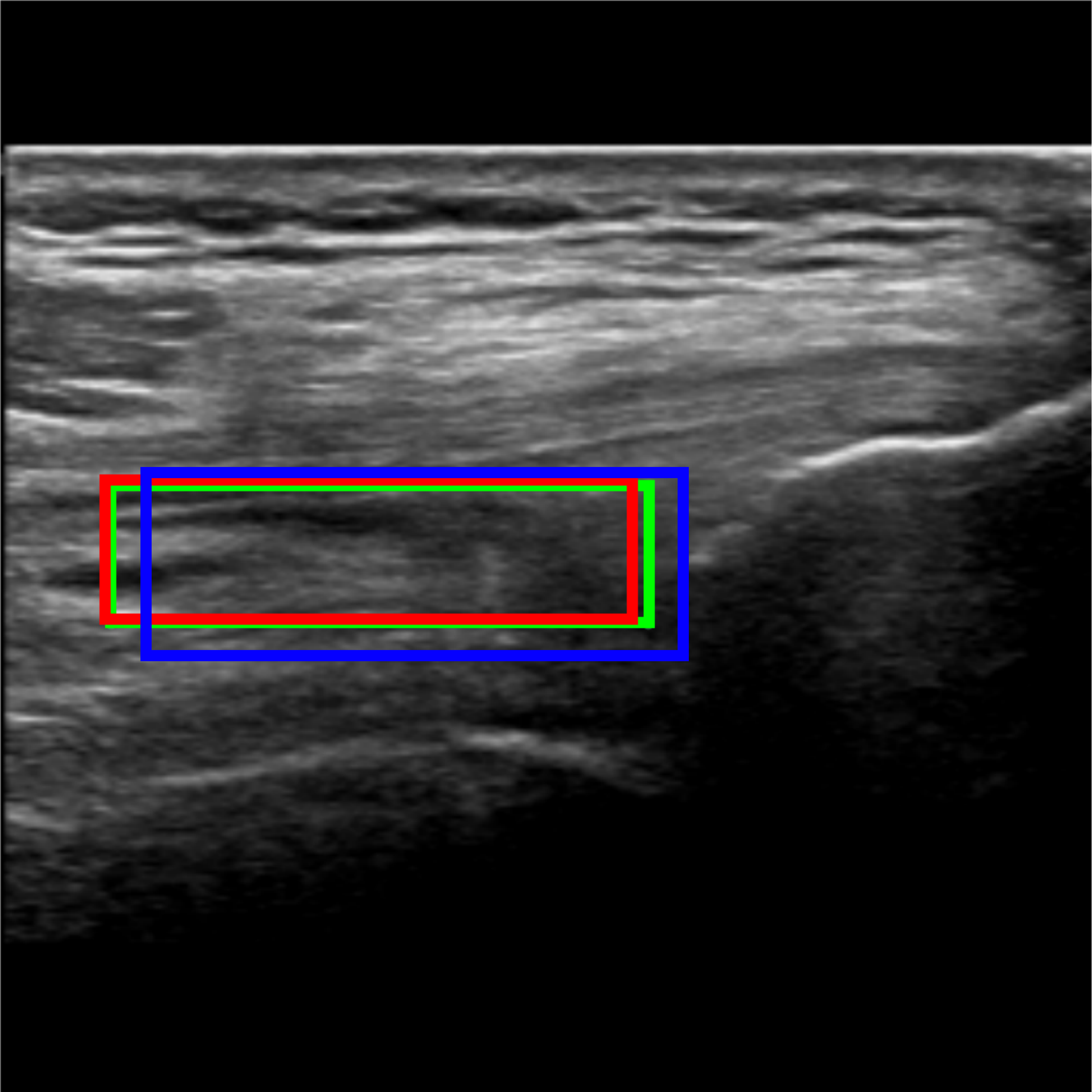}}
\\
\subfloat[Worst detection by \textit{Detection approach}\label{fig:min_iou_det}]{
\includegraphics[width=.4\textwidth]{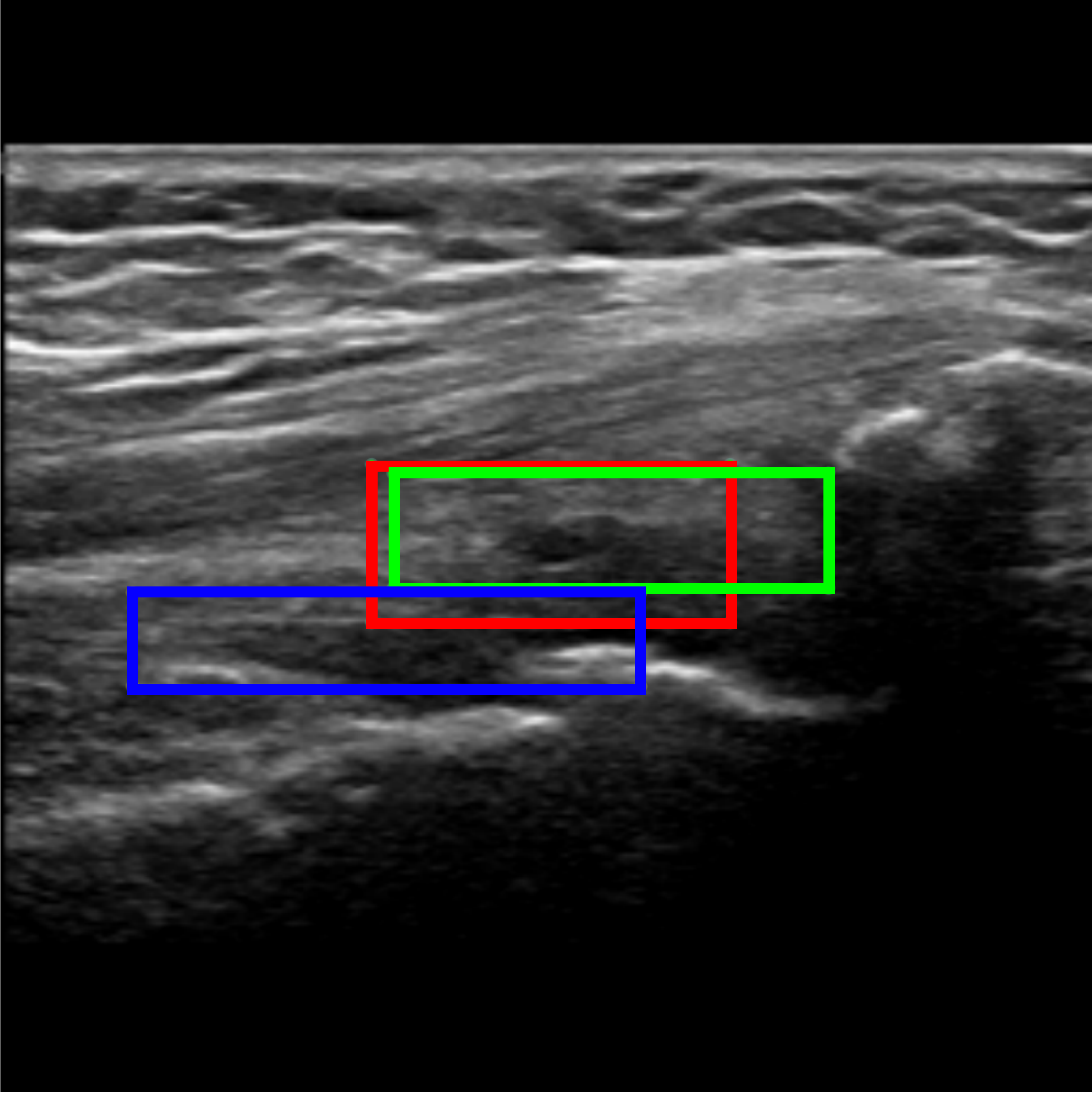}}
\hfill
\subfloat[Best detection by \textit{Detection approach} \label{fig:max_iou_det}]{
\includegraphics[width=.4\textwidth]{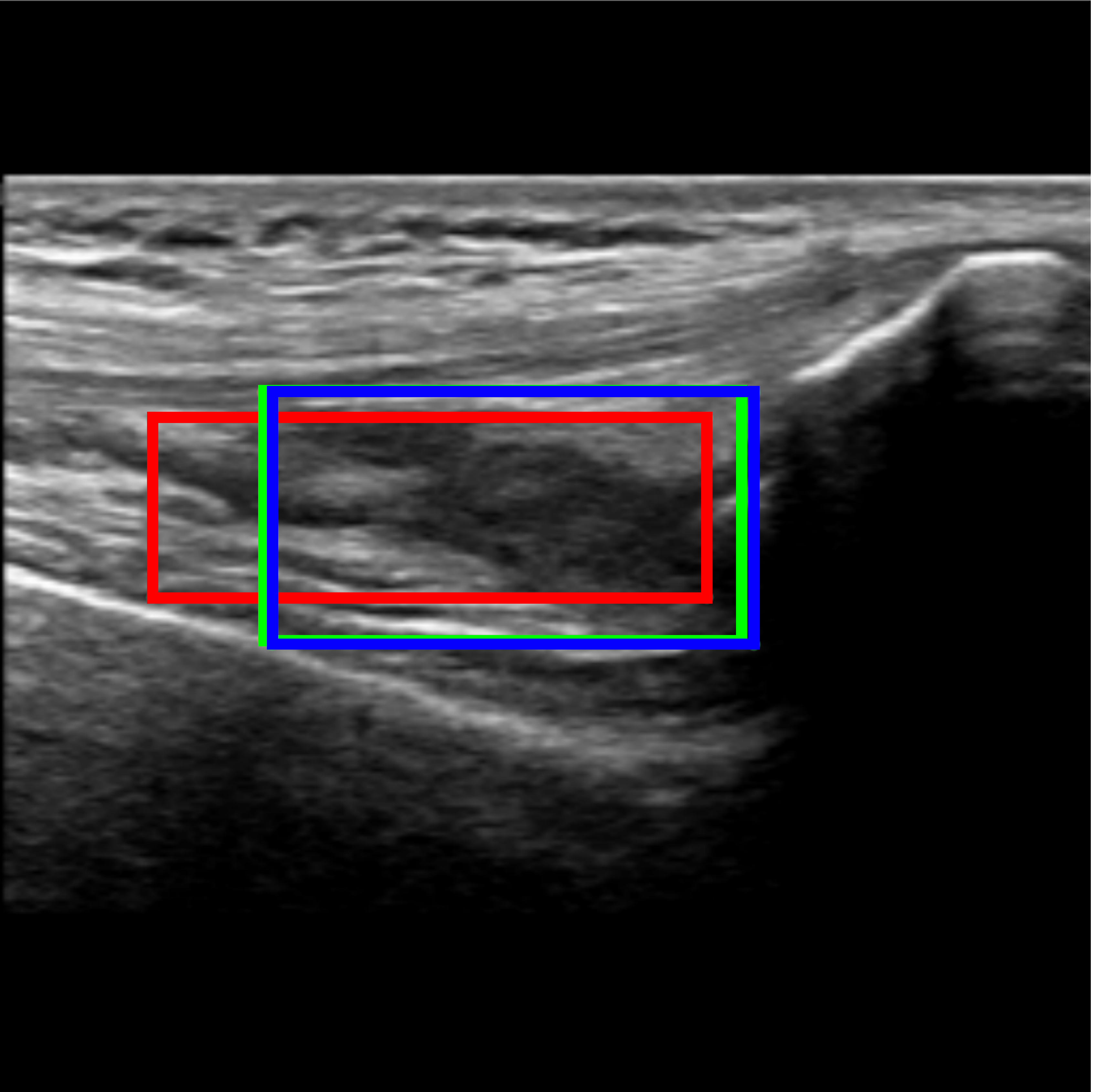}}
\caption{Detection examples. Green represents the ground truth, red and blue the results of the \textit{Multi-Task approach} and \textit{Detection approach}, respectively.}
\label{fig:multi_iou}
\end{figure}

Considering the detection problem, Figure~\ref{fig:multi_iou} shows US images where the two approaches detected the SQR with the lowest and the highest IoU.
In Figure~\ref{fig:min_iou_multi}, the \textit{Multi-task approach} wrongly detects as SQR an image region that is similar to an actual SQR in terms of position and shape, resulting in a very low value of IoU ($0.33$). In this case, also the \textit{Detection approach} can not reliably detect the right target precisely, and indeed it  detects only a small portion of the actual SQR (IoU=$0.33$).
Instead, in the example shown in Figure~\ref{fig:max_iou_multi} the \textit{Multi-task approach} accurately detects the SQR (IoU=$0.95$), while the \textit{Detection approach} identifies the same area with a lower IoU ($0.68)$.

Figure~\ref{fig:min_iou_det} shows the US image for which the \textit{Detection approach} provided the lowest IoU value. The problem is similar to that of Figure~\ref{fig:min_iou_multi}: a region is erroneously recognized as a SQR because it is similar to a SQR.
In this case the detected bounding box does not overlap with the ground truth, hence the IoU is zero. Instead, the \textit{Multi-Task approach} basically detects the right target (IOU=$0.58$).

Figure~\ref{fig:max_iou_det} shows instead the US image for which the \textit{Detection approach} provided the highest IoU value ($0.96$). In this case the \textit{Multi-task approach} identifies the right target less precisely, resulting in a IoU of $0.55$.
\section{Conclusion}
Early detection of hemarthrosis is fundamental to reduce the risk of under and over treatment for hemophilic patients. A Computer-Aided Diagnosis (CAD) tool that detects joint recess distension from ultrasound (US) images can support practitioners in diagnosing hemarthrosis without the need for expensive and time consuming exams (like MRI).
We investigate the requirements of such a tool and we frame the problem in terms of a combination of two typical machine learning tasks: classification and detection.
Addressing this problem is particularly challenging for a number of reasons, including that the position and the shape of the joint recess may change considerably across different US images, and there can even be borderline cases in which the recess is only partially distended.
Finally, the datasets in this problem domain are generally small and may contain noisy images. 

This paper presents two solutions, each providing both recess detection and classification. Experiments, conducted on images of the Subquadricipital Recess (SQR), \textit{i.e.}, the knee joint recess, focusing on a specific US scan (the SQR longitudinal scan) show promising results. Indeed, both solutions achieve a Balanced Accuracy (BA) of approximately $0.75$, a threshold value used in the literature to distinguish ``useful'' medical tests~\cite{power2013principles}. In particular the \textit{MultiTask approach} achieves a BA value of $0.78$.
For what concerns the detection, the two solutions guarantee a correct detection (\textit{i.e.}, IoU $> 0.5$) in more then $82\%$ of the cases.

We believe that the performance of our solutions can be considerably improved in two possible ways. First, the CAD tool could process multiple images of the same joint, possible from different scans or from a video feed. The different results computed on the various images can then be combined to provide a more reliable outcome. The second improvement could be the adoption of an ensemble approach, in which a number of different models are trained and the CAD tool computes the result as a combination of the individual results provided by each model. Beyond improving the performance, these possible improvements have a potential important advantage: they can identify borderline cases (e.g., when there is a disagreement in classification by two or more models, or by processing two scans of the same knee). In these (hopefully rare) cases, the CAD tool can inform the practitioner who can decide, for example, to use a different diagnosing tool (e.g., MRI).

As a future work we intend to apply the proposed solutions on multiple scans for multiple joints. Actually, we are currently collecting images and videos on a total of $6$ scans for the knee, the elbow, and the ankle.
Our final goal is to create a bed-side solution for early hemarthrosis diagnosis. The idea is to enable an operator with little training (\textit{e.g.}, the patient or a caregiver) to acquire US images with a portable US device. The system will support the operator during the acquisition by identifying the relevant reference points (\textit{e.g.}, the patella), by guiding the operator to correctly position the US probe, and by evaluating the images in real time to inform the operator that a suitable image was collected. The US images will then be transmitted to the practitioner or even automatically processed as a part of a screening procedure.
%


\section*{Acknowledgements}
This work was partially supported by the project ''MUSA - Multilayered Urban Sustainability Action'',  NextGeneration EU, MUR PNRR.

\bibliographystyle{unsrt}
\bibliography{biblio}  






\end{document}